\documentclass[preprint,showpacs,preprintnumbers,amsmath,amssymb,nofootinbib]{revtex4-1}
\usepackage[dvips]{graphicx}
\usepackage{graphicx}
\usepackage{amsfonts}
\usepackage{bm}
\usepackage{amsmath}
\usepackage{amssymb}
\usepackage{color}
\usepackage[all]{xy}

\def\be{\begin{equation}}
\def\ee{\end{equation}}
\def\bea{\begin{eqnarray}}
\def\eea{\end{eqnarray}}

\begin{document}

\title{Thermodynamics of the FLRW apparent horizon}

\author{Luis M. S\'anchez$^1$ and Hernando Quevedo$^{1,2}$   }
\email{luis.sanchez@correo.nucleares.unam.mx,quevedo@nucleares.unam.mx}
\affiliation{$^1$Instituto de Ciencias Nucleares, 
	Universidad Nacional Aut\'onoma de M\'exico, 
	AP 70543, M\'exico, DF 04510, Mexico}
\affiliation{$^2$ 
	Dipartimento di Fisica and ICRA, Universit\`a di Roma ``La Sapienza", Piazzale Aldo Moro 5, I-00185 Roma, Italy}

\begin{abstract}
To investigate the relationship between gravity and thermodynamics in the case of dynamic systems, we interpret the apparent horizon of the Friedmann-Lema\^itre-Robertson-Walker (FLRW) spacetime as a thermodynamic system. We derive the corresponding fundamental equation in the entropy representation and explore the consequences of demanding the fulfillment of the classical laws of thermodynamics. 
We investigate in detail the case of an FLRW spacetime with and without cosmological constant. We show that the interpretation of the apparent horizon as a thermodynamic system is possible only  under certain conditions because the analogy is not always valid for all the values of the  parameters entering the dynamics of the horizon and certainly not for its entire evolution.

\end{abstract}


\maketitle

\section{Introduction}
\label{sec:int}

It is well known that the laws of black hole  mechanics show a remarkable similarity with the laws of classical thermodynamics (for a recent review, see, for instance,  \cite{wald2001thermodynamics} ) Although the physical explanation of this similarity is still an open problem, it seems to indicate that there exists a deep link between gravity and thermodynamics. One of the main features of this analogy consists in identifying the entropy with the area of an event horizon. 

However, already at the level  of a flat spacetime the Unruh effect points out to a similar relationship between temperature and acceleration. In fact, the temperature is ascribed to a null surface that acts as a horizon for observers moving along accelerated trajectories \cite{unruh1976notes}.

In the above examples, the horizon plays an important role as the surface to which thermodynamic properties are associated. However, an example in which no horizon is needed for establishing the link between gravity and thermodynamics is the case of a spherically symmetric perfect fluid in equilibrium, where the Einstein equations emerge as a result of demanding that the total entropy of the matter be an extremal \cite{oppenheim2001thermodynamic}.

A further indication of the  inter-relationship between gravity and thermodynamics was the proof that Einstein's equations can be interpreted as a first law of thermodynamics \cite{jacobson1995thermodynamics}.
Further studies have shown that this relationship  holds also in other gravity theories (for a review, see \cite{padmanabhan2014general}). Moreover, in the case of dynamic cosmological models similar analogies have been established. In fact, by applying the first law of thermodynamics to the apparent horizon of a homogeneous and isotropic universe in Einstein theory and by assuming that the entropy is related to the area of the apparent horizon, the dynamic Friedmann equations were derived \cite{cai2005first}.

In this work, we continue the investigation of the link between gravity and thermodynamics in the case of cosmological spacetimes. We will assume that the FLRW apparent horizon can be interpreted as a thermodynamic system. According to the standard approach of classical thermodynamics \cite{callen1998thermodynamics}, this implies that there should exist a fundamental  equation  from which all the thermodynamic  properties of the horizon can be derived. In fact, we use the definition of the apparent horizon radius and Friedmann equations to derive the corresponding fundamental equation, which is then investigated in detail 
by considering the first and second laws of thermodynamics and the corresponding response functions. This procedure is performed in the case of an FLRW spacetime with and without cosmological constant. We will show that the second law is not always satisfied, implying that in certain cases the apparent horizon cannot be interpreted as a thermodynamic system.

This work is organized as follows. In Sec. \ref{sec:review}, we review the main aspects of the FLRW spacetime within Einstein's theory of gravity with a perfect-fluid source. The link between gravity and thermodynamics in the case of apparent horizons is also reviewed. Then, in Sec. \ref{sec:flrw}, we derive the main properties of the FLRW apparent horizon and analyze its dynamic behavior. The fundamental equation and the thermodynamic properties of an apparent horizon, which satisfies a barotropic equation of state, are presented in Sec. \ref{sec:feq} by applying the laws of thermodynamics and by computing the response functions. A similar approach is applied to the case of the apparent horizon of an FLRW spacetime with cosmological constant in Sec. \ref{sec:lambda}. Finally, the conclusions are drawn in Sec. \ref{sec:con}. Throughout this work we used geometric units with $G=1$, $c=1$, $k_{_B}=1$, and $\hbar =1$.

\section{FLRW cosmology and thermodynamics}
\label{sec:review}

We will consider an FLRW universe with line element $(k=0,\pm 1)$
\be
ds^2 = -dt^2 + a^2(t)\left[ \frac{dr^2}{1-kr^2} + r^2 (d\theta^2 + \sin^2\theta d\varphi^2)\right]\ ,
\ee
which for a perfect fluid with energy-momentum tensor, $T_{\mu\nu} = (\rho+p)u_\mu u_\nu + p g_{\mu\nu}$, leads to the general relativistic Friedmann equations 
\be
H^2+ \frac{k}{a^2} = \frac{8\pi }{3}\rho\ ,
\label{fried1}
\ee
\be
\dot H - \frac{k}{a^2} = -4\pi(\rho+p) \ ,
\label{fried2}
\ee
where $H=\frac{\dot a}{a}$ is the Hubble parameter. 
It is convenient to consider also the energy-momentum conservation law
\be
\dot \rho + 3H(\rho+p)=0\ ,
\label{claw}
\ee
which follows from the Friedmann equations.
For concreteness, let us consider a barotropic fluid with $p=w \rho$, where $w$ is the constant barotropic factor. Then, from the conservation law (\ref{claw}), we obtain that 
\be
\rho = \rho_0 a^{-3(1+w)}\ ,
\label{rho}
\ee 
where $\rho_0$ is an integration constant.  Instead of the scale factor, we can also use the definition of the redshift, $z= \frac{a_0}{a} - 1$, where $a_0$ is a constant,
to re-express all the quantities in terms of $z$. So, the evolution of the density function can be described by means of the equation
\be
\rho = \rho_0 \left(\frac{1+z}{a_0}\right)^{3(1+w)}\ .
\label{rho2}
\ee

On the other hand, in the case of a dynamical system such as a cosmological model, the laws of thermodynamics must be re-expressed in such a way that they are valid at each moment of the evolution of the system. In the case of spherically symmetric apparent horizons, it has been established that the first law can be expressed as 
\be
dE = A \Psi + W dV 
\label{uflaw1}
\ee
and is dubbed unified first law \cite{hayward1998unified,hayward1994general,hayward1996gravitational}.  Here $A$ is the surface of a sphere of radius $R$, $\Psi$ is the energy flux 
\be
\Psi = \Psi_ a dx^a = \left(T_a^{\ b}\frac{\partial R}{\partial x^b} + W \frac{\partial R}{\partial x^a}\right)dx^a \ ,
\ee
and $W$ is the work density
\be
W = - \frac{1}{2} T^{ab}h_{ab}\ .
\ee
In the case of the FLRW spacetime with a perfect fluid, we obtain at the apparent horizon
\be
dE =  2{\pi} R_h^2 (\rho+p)\left(-HR_h dt + a dr\right)+\frac{1}{2} (\rho-p)d V\ ,
\label{uflaw2}
\ee
and 
\be
W=\frac{1}{2} (\rho-p)\ .
\ee


\section{The FLRW apparent horizon }
\label{sec:flrw}

The homogeneous and isotropic FLRW line element can be represented as \cite{cai2005first,zhang2021thermodynamics}
\be
ds^2 = h_{ab}dx^a dx^b + R^2 (d\theta^2 +\sin^2 \theta d\varphi^2)\ ,
\ee
where $x^0 = t$, $x^1 =r$, $R=a(t) r$, and
\be
h_{ab}= {\rm diag}\left(-1,\frac{a^2(t)}{1-kr^2}\right) .
\ee
A particular property of this spacetime is that it possesses a marginally trapped null surface with vanishing expansion, the apparent horizon, that  satisfies the relationship
\be
h^{ab}\frac{\partial R}{\partial x^a}\frac{\partial R}{\partial x^b} =0\ .
\ee
The solution of this equation 
\be
R_h = \frac{1}{\sqrt{
		H^2 + \frac{k}{a^2} } }
\label{hor}
\ee
determines the apparent horizon as a dynamic structure that evolves in time.

Furthermore, it has been argued that the apparent horizon of a dynamic system can be interpreted as a causal horizon, to which we can associate a surface gravity  and a gravitational entropy  \cite{hayward1994general, hayward1996gravitational,bak2000cosmic,hayward1998unified,faraoni2011cosmological,melia2018apparent}.  
In the case of an FLRW universe, the surface gravity of the apparent horizon is defined as \cite{cai2005first,hayward1998unified, binetruy2015apparent}
\begin{equation}
\kappa = \frac{1}{2 \sqrt{-h}} \frac{\partial}{\partial x^{a}} \left(\sqrt{-h} h^{ab} \frac{\partial R_h}{\partial x^{b}} \right)
\end{equation}
where $h$ is the determinant of $h_{ab}$. Then, we obtain
\be
\kappa = -\frac{1}{R_h } \left(1- \frac{1}{2} 
\frac{ \dot{R}_h}{ H R_h } \right)\ .
\ee
Furthermore, considering the apparent horizon as a thermodynamic system, we can define  the temperature   as $T_h= \frac{\vert\kappa\vert}{2\pi}$, which leads to 
\begin{equation}
T_h = \frac{1}{2 \pi R_h } \left|1- \frac{1}{2}\frac{ \dot{R}_h}{ H R_h} \right|\ .
\label{temp}
\end{equation}
Notice that the temperature depends not only on the value of the horizon radius, but also on its change along the evolution, i. e., on $\dot R_h$. This is a consequence of the dynamic character of the apparent horizon. 
Notice also that in the limiting case of a very slowly changing apparent horizon or for $\frac{\dot{R}_h}{2 H R_h} \ll 1$, the temperature reduces to $
T_h=\frac{1}{2\pi R_h}$,
an expression that resembles the temperature of a spherically symmetric black hole with horizon radius $R_h$.
To investigate the dynamic behavior of the temperature, we calculate the rate of change  of the horizon radius, i.e.,
\begin{equation}
\dot{R}_h = -R^{3}_h H \left( \dot{H}-\frac{k}{a^{2}} \right) = 4\pi R_h^3 H (\rho + p)\ ,
\label{dotR}
\end{equation}
where we have used the second Friedmann equation (\ref{fried2}). Then, from Eq.(\ref{temp}) we obtain
\be
T_h = \frac{1}{2\pi R_h}\left|1 - 2\pi R_h^2(\rho+p)\right|\ ,
\label{temp2}
\ee
which is the most general expression for the temperature of the FLRW apparent horizon. 

In addition, we introduce the dynamic entropy
\be
S= \frac{1}{4} A = \pi R_h^2 = \pi \left(H^2 + \frac{k}{a^2}\right)^{-1}\ ,
\label{ent0}
\ee 
where $A$ is the area of the apparent horizon. As a function of time, the entropy should satisfy the conditions $\dot S \geq 0$ and $\ddot S \leq 0$,   
which mean that the entropy should be an increasing function that fulfills the principle of maximum entropy.
In the next section, we will analyze these conditions in detail.

\begin{figure}
	\includegraphics[scale=0.35]{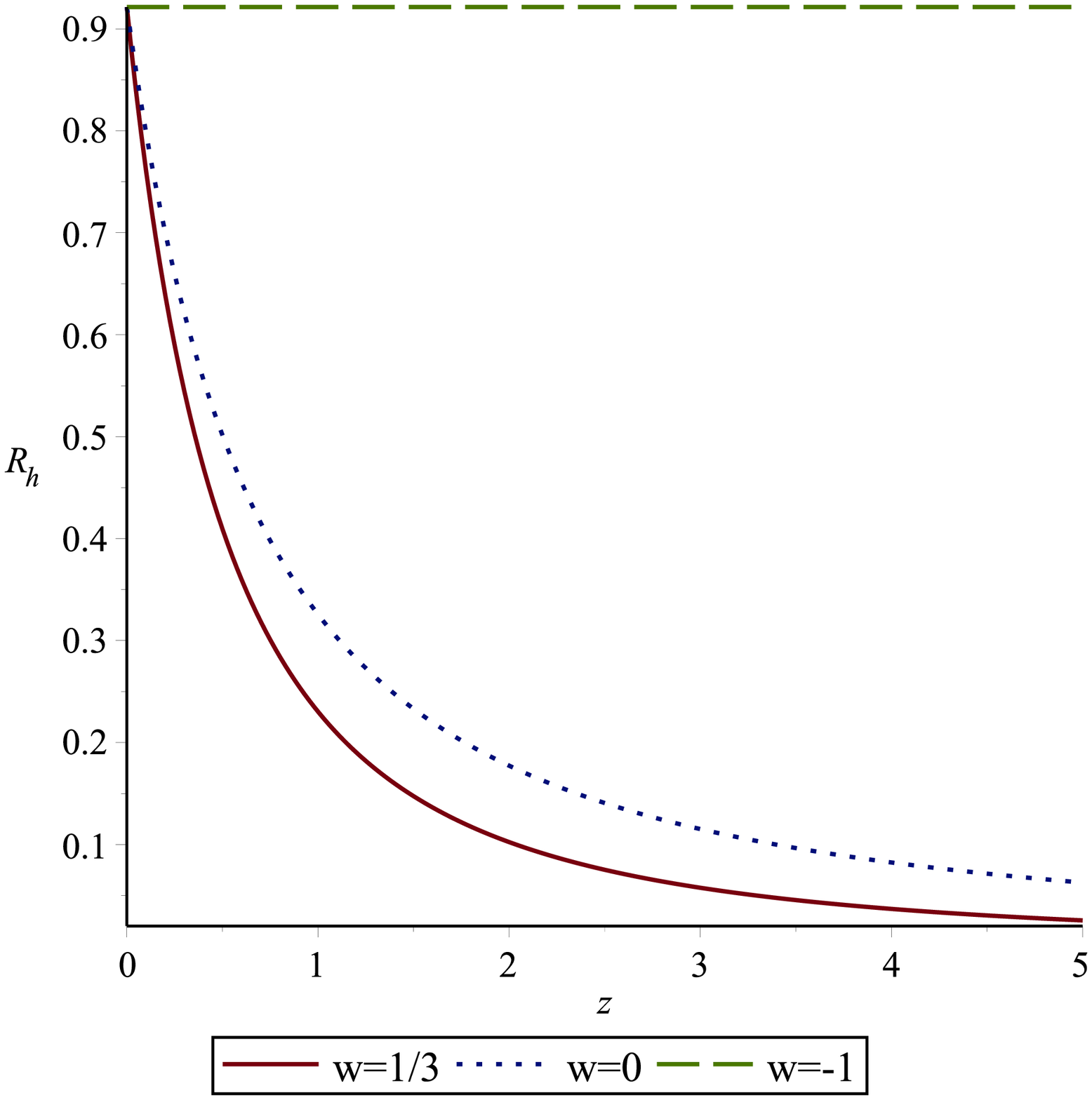} \qquad
	\includegraphics[scale=0.35]{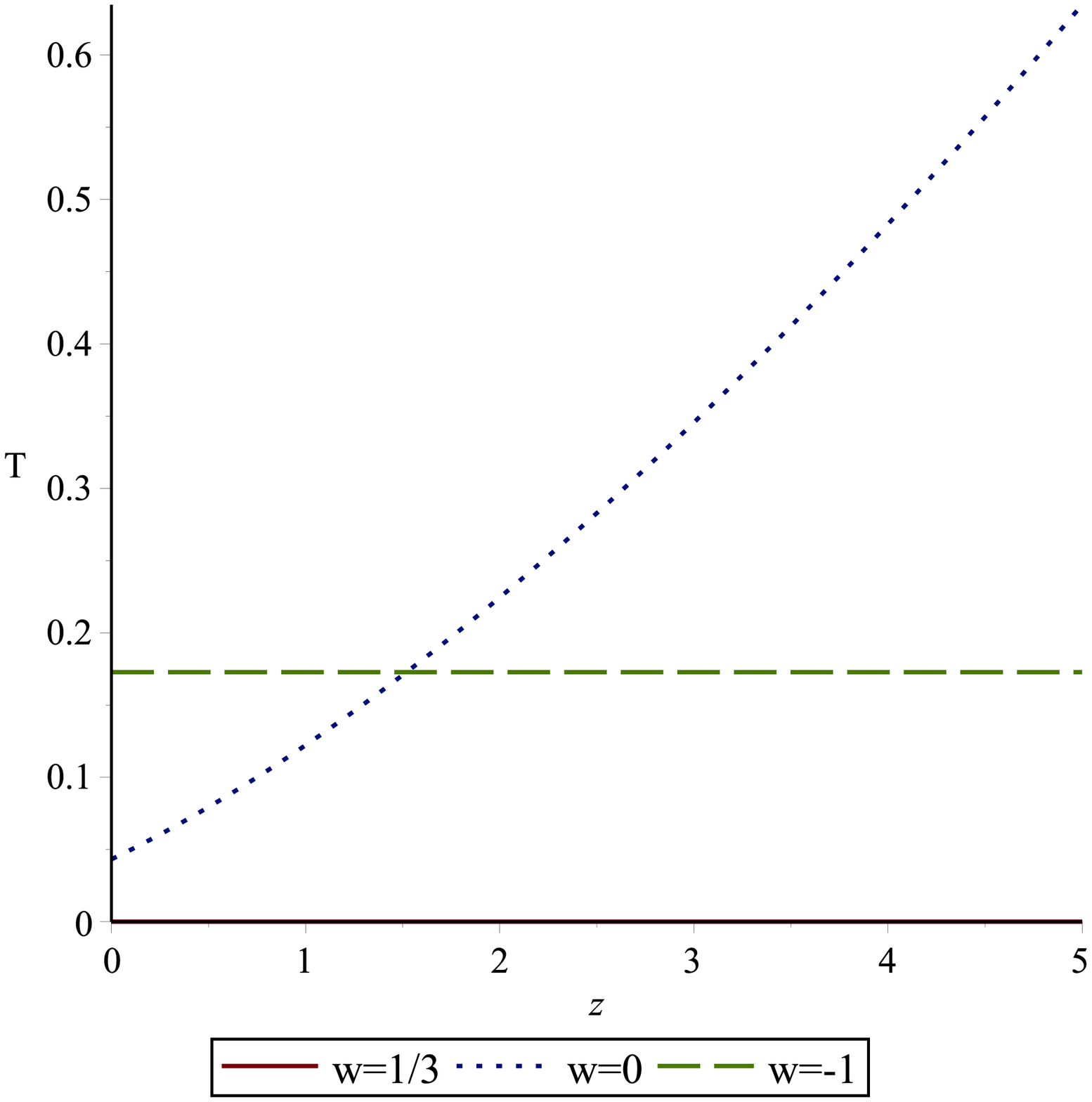}
	\includegraphics[scale=0.35]{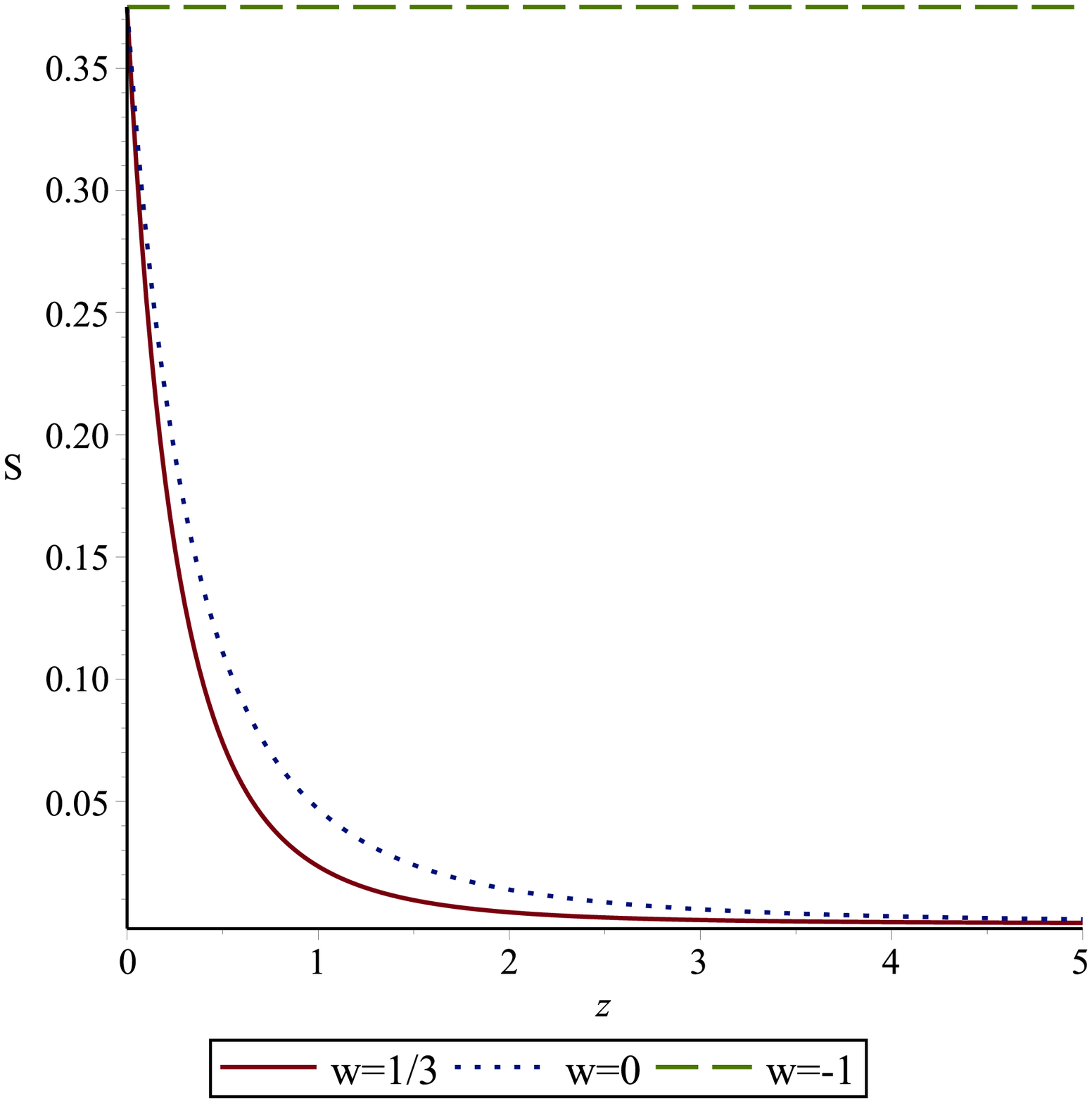}
	\caption{Radius, temperature and entropy of the apparent horizon in terms of the redshift parameter for different values of the barotropic constant $w$. Here, we choose $\rho_0=1$ and $a_0=1$ so that present time corresponds to $z=0$.}
	\label{fig1}
\end{figure}

The dynamic behavior of the apparent horizon and its main thermodynamic variables can be analyzed re-expressing them in terms of the energy density $\rho$ by using the first Friedmann equation.  
Then, from Eqs.(\ref{rho2}), (\ref{hor}), (\ref{temp2}) and (\ref{ent0}), we obtain the expressions
\be
p=w\rho_0 \left(\frac{1+z}{a_0}\right)^{3(1+w)}\ ,
\ee
\be
R_h = \sqrt{\frac{3}{8\pi\rho_0}}\, \left(\frac{1+z}{a_0}\right)^
{-\frac{3}{2}(1+w)}\ ,
\ee
\be
T_h =  \sqrt{\frac{2\rho_0}{3\pi}}\, \left|1-\frac{3}{4}(1+w)\right|\,
\left(\frac{1+z}{a_0}\right)^{\frac{3}{2}(1+w)}\ ,
\ee
\be
S = \frac{3}{8\rho_0}\left(\frac{1+z}{a_0}\right)^{-3(1+w)}\ ,
\ee
where we have used the expression (\ref{rho2}) for a barotropic density in terms of the redshift.
In Fig. \ref{fig1}, we illustrate the behavior of the  radius, temperature, and entropy of the horizon for different values of the barotropic factor. 
Notice that in the case of a dark energy universe $(w=-1)$, the radius, temperature, and entropy are constant, i.e., the apparent horizon does not evolve in time. In this case, the entropy condition (\ref{entcon1})
and (\ref{entcon2}) are trivially satisfied.
In the case of a matter and radiation dominated universe with $w=0$ and $w=1/3$, respectively, the radius of the horizon increases as the redshift diminishes, reaching its maximum value for $z=0$. Moreover, the temperature of a matter universe decreases with the redshift. Notice that in the case of a radiation dominated universe, the temperature of the horizon vanishes at all times, indicating a limiting value for the application of the thermodynamic analogy.


\section{Fundamental equation and thermodynamic properties}
\label{sec:feq}

The apparent horizon of the FLRW spacetime is a dynamical quantity that depends explicitly on the value of the  scale factor $a(t)$, as shown in Eq.(\ref{hor}). On the other hand, the scale factor satisfies the first Friedmann equation (\ref{fried1}). Combining these two expressions, we obtain 
\be
S = \frac{3}{8\rho} = \frac{3 V}{8E} \ ,
\label{ent}
\ee
where $E$ is the total energy contained inside a sphere of radius $R_h$. This function  relates extensive thermodynamic variables and can be considered as the fundamental equation of the corresponding thermodynamic system in the entropic representation \cite{callen1998thermodynamics}. 
Then, as in classical thermodynamics, the fundamental equation should satisfy the laws of thermodynamics and contain the entire thermodynamic information of the system.

\subsection{The first law}
\label{sec:flaw}

From the fundamental equation (\ref{ent}), we can derive the conservation of energy condition
\be
dE = - \frac{3V}{8S^2}dS + \frac{3}{8S}dV = -\frac{8}{3}V\rho^2 d S + \rho dV\ ,
\label{enc}
\ee
which should be related to the first law of thermodynamics, i.e., $\delta Q = dU +  W dV $, where $\delta Q$ is the amount of heat absorbed by the system, $dU$ is the change of internal energy, and $ W dV $ is the work done by the system. In the case of the apparent horizon, the change in internal energy $dU$ is just the amount of energy crossing it during the evolution of the horizon, i.e., $dU=-dE$ \cite{cai2005first}. 
Then, we obtain $dE=-\delta Q + W dV $. Moreover, the change in entropy $dS$ is related to the amount of absorbed heat as $dS = \delta Q/T$, where $T$ is the temperature. Consequently, the first law can be expressed as 
\be 
dE = - TdS + W dV
\label{flaw}
\ee
and from Eq.(\ref{enc}), we obtain
\be
T =  \frac{8}{3}V \rho^2 \ ,\quad  W = \rho \ .
\label{tw}
\ee
These are the main thermodynamic variables  that are obtained for the apparent horizon from the fundamental equation (\ref{ent}).

On the other hand, as mentioned in Sec. \ref{sec:int}, from the unified first law for the FLRW apparent horizon (\ref{uflaw2}), it follows that in general the work density $W$ should be \cite{cai2005first} 
\be 
W=\frac{1}{2} (\rho-p)\ .
\ee
Comparing this formula with Eq.(\ref{tw}), which was obtained from the fundamental equation (\ref{ent}), we conclude that both expressions coincide only for 
\be
\rho + p =0\ ,
\label{eos1}
\ee
which represents a barotropic equation of state with barotropic factor $w=-1$. Moreover, it follows  from Eq.(\ref{dotR}) that $\dot R_h =0$. This means that if we associate the entropy $S = \frac{A}{4}$ to the apparent horizon and consider it as the fundamental equation, the corresponding thermodynamic system corresponds to that of a dynamic apparent horizon with temperature $T= (2\pi R_h)^{-1}$. This particular case of a dynamic apparent horizon has also been investigated in \cite{cai2005first} \cite{debnath2020thermodynamics} just by assuming  that $\frac{\dot{R}_h}{2 H R_h} \ll 1$ in Eq.(\ref{temp}). Here, we have found the fundamental equation for this particular case. 
The advantage of using a fundamental equation is that we can derive all the thermodynamic properties of the system. Indeed, from the above discussion we conclude that in this case the main thermodynamic variables
are
\be
S = \frac{3}{8\rho}\ , \ T = \frac{8}{3} V\rho^2\ , \ p=-\rho\ ,
\label{main1}
\ee
which, by using the  solution of the conservation law (\ref{claw}) and the definition of the redshift parameter, for a barotropic perfect fluid with $w=-1$ can be expressed as
\be
S = \frac{3}{8\rho_0}\ , \ T = 
\sqrt{\frac{2\rho_0}{3\pi}} 
\ , \ p=-\rho_0\ ,
\label{main2}
\ee
so that all the main thermodynamic quantities remain constant.

\subsection{The second law}
\label{sec:slaw}

The entropy function should satisfy the condition
\be
\dot S =-\frac{3\dot\rho}{8\rho^2}  
  \geq 0 \ ,
\label{entcon1}
\ee
which implies that the density should be a decreasing function of time, i.e.,  $\dot\rho \leq 0$.  According to the conservation law (\ref{claw}) for a barotropic fluid, which leads to
\be 
\dot S 
= \frac{9H}{8\rho}(1+w)\geq 0 \ ,
\ee
for  positive Hubble parameter and energy density, the first entropy condition  implies that $w\geq -1$.
 
\begin{figure}
	\includegraphics[scale=0.35]{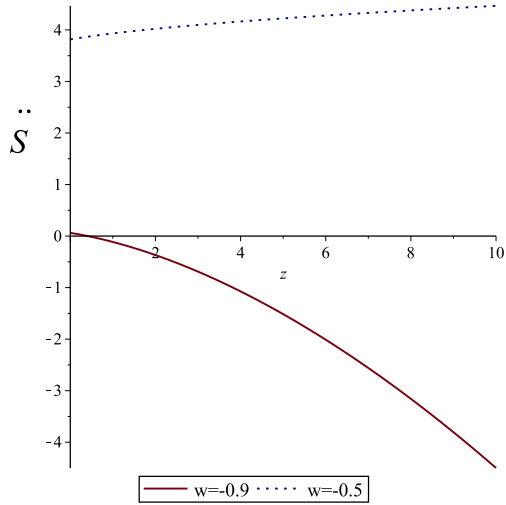}
	\caption{Second derivative of the entropy as a function of the redshift for different values of the barotropic parameter. The constants are chosen as $\rho_0=1$, $a_0=1$, and $k=-1$. }
	\label{fig2}
\end{figure}

Moreover, the condition $\ddot S\leq 0$ leads to
\bea
\ddot S &=& \frac{3}{8}\left(2\frac{\dot\rho^2}{\rho^3}-
\frac{\ddot \rho}{\rho^2}\right) \nonumber\\
&=& \frac{9}{2}\pi (1+w)^2 - \frac{9k}{8\rho_0}(1+w)(2+3w)\left(\frac{1+z}{a_0}\right)^{-1-3w}\leq 0\  ,
\label{entcon2}
\eea
where we have used the density function (\ref{rho}) and the Friedmann equations to express the derivatives of the scale factor in terms of the density. In the case $k=0$, the only possible solution is $w=-1$, which corresponds to a ``dark energy" fluid. This solution is also valid in the case $k\neq 0$. In general, since the function $(1+z)^{-1-3w}$ is positive, the condition $\ddot S< 0$ can be satisfied only within the range of values of $k$ and $w$ that fulfills  the condition $k(1+w)(2+3w)>0$. For $k=1$, this means that only barotropic fluids with $w<-1$ or $w>-2/3$ are allowed. Furthermore, for $k=-1$, only fluids with barotropic factor $w\in (-1,-2/3)$ can be considered as thermodynamic systems that represent apparent horizons.
We illustrate this behavior in Fig. \ref{fig2}. We see that 
the fluid with $w=-1$ corresponds to an entropy function with $\ddot S=0$  that trivially satisfies the entropy condition. The fluid with $k=-1$ and $w=-0.5$ is not within the allowed range and so the entropy can never satisfy the condition $\ddot S \leq 0$, 
indicating that the apparent horizon in an universe with $k=-1$ cannot be interpreted as a thermodynamic system corresponding to a barotropic fluid with $w=-0.5$.  
On the contrary, the fluid with $k=-1$ and $w=-0,9$ is within the allowed range and, therefore, it satisfies the condition $\ddot S\leq 0$ most of the time. 
In fact, as can be seen from Fig. \ref{fig2}, at late times this condition is no more valid due to the fact that the second term on the right-hand side of Eq.(\ref{entcon2})  depends explicitly on the redshift parameter.

Taking into account also the main consequence of the first law, which is contained in Eq.(\ref{eos1}), we conclude that the FLRW apparent horizon can be interpreted only as barotropic fluid with $w=-1$, i.e., as a "dark-energy" fluid.

\subsection{The response functions}
\label{sec:resp}

To further analyze the thermodynamic properties of the apparent horizon, we calculate the heat capacities
\be
C_V = T\left(\frac{\partial S}{\partial T}\right)_V\ , \quad
C_p = T\left(\frac{\partial S}{\partial T}\right)_p \ ,
\ee
the compressibilities
\be
\kappa_T = - \frac{1}{V} \left(\frac{\partial V}{\partial p}\right)_T\ , \quad
\kappa_S = - \frac{1}{V} \left(\frac{\partial V}{\partial p}\right)_S\ , 
\ee
and the thermal coefficient
\be
\alpha= \frac{1}{V}\left(\frac{\partial V}{\partial T}\right)_p\ .
\ee
To perform these computations, we use the relationships
\be
S = \sqrt{\frac{3V}{8T}}\ , \quad
V= \frac{3T}{8p^2}\ ,
\ee
which follow from Eq.(\ref{main1}). Then, we obtain
\be
C_V = -\frac{1}{2}\sqrt{\frac{3V}{8T}}\ , \quad \kappa_T = \frac{3}{4}\frac{T}{Vp^3} \ ,\quad 
\alpha = \frac{3}{8Vp^2} \ .
\label{resp1}
\ee

Furthermore, we use the relationships \cite{callen1998thermodynamics}
\be
C_p=C_V+ \frac{TV \alpha^{2}}{ \kappa_T}\ , \qquad \kappa_S=\kappa_T- \frac{TV \alpha^{2}}{ C_p}\ ,
\ee
to obtain
\be
C_p = - \sqrt{\frac{3V}{8T}},\quad \kappa_S = \frac{3}{8}\frac{T}{Vp^3}\ .
\ee
Finally, the adiabatic index $\gamma$ can be determined as 
\be
\gamma= \frac{C_p}{C_V} = \frac{\kappa_T}{\kappa_S}  =  2\ .
\ee
Some interesting features about the thermodynamic properties of the apparent horizon follow from the analysis of the above expressions. First of all, we notice that the heat capacities $C_V$ and $C_p$ are negative, a property which could be interpreted as contradictory in the case of ordinary thermodynamic systems. However, it is well known that long-range interactions can give rise to negative heat capacities
\cite{lynden1968gravo,lynden1980consequences, goodman2012dynamics,escamilla2019statistical}.
 In the case of apparent horizons, the dynamical behavior is governed by the long-range gravitational interaction, which explains the negative sign in front of the heat capacities.  

To further analyze the response functions given above, we consider their explicit dependence on the redshift parameter $z$. Replacing the expressions for the main thermodynamic variables (\ref{main1}) and (\ref{main2}) into the expressions for the response functions (\ref{resp1}), we obtain
\be
C_V = - \frac{3}{16\rho_0}\ , \quad \kappa_T = - \frac{2}{\rho_0}\ , \quad \alpha=
\sqrt{\frac{3\pi}{2\rho_0}}
\ .
\ee
We conclude that all the response functions in this case are constant. This result is physically compatible with the results obtained for the main thermodynamic variables in Eq.(\ref{main2}), which indicate that all of them are constant. Moreover, we see that the compressibilities are also negative, which can be explained by the long-range character of the gravitational interaction. 

The fact that the heat capacities and compressibilities are constants during the evolution of the apparent horizon allows us to perform a formal comparison with the properties of an ideal gas. In fact, in an ideal gas such response functions are positive constants. In the case of an apparent horizon the constants are negative, as mentioned above, due to the long-range gravitational interaction. However, the corresponding adiabatic index $\gamma$ is positive as in the case of ideal gases. We can, therefore, establish a formal analogy with the expression for the adiabatic index of an ideal gas \cite{callen1998thermodynamics} \cite{huang2009introduction}

\be
\gamma = \frac{2+f}{f}\ ,
\ee
where $f$ is the number of degrees of freedom of the system. Then, we have that the apparent horizon as a thermodynamic system possesses $f=2$ degrees. This contracts with the simplest case of a monoatomic gas in which $f=3$.  
Probably, this reduction of the numbers of degrees of freedom is a consequence of the presence of the long-range gravitational field.


\section{The FLRW apparent horizon with cosmological constant} 
\label{sec:lambda}

Consider a FLRW spacetime with cosmological constant $\Lambda$. In this case, the corresponding Friedmann equations can be written as
\be
\frac{\dot a^2}{a^2} + \frac{k}{a^2} = \frac{8\pi}{3} (\rho - P)
\ ,
\label{fl1}
\ee
\be   \frac{\ddot a}{a} = -\frac{4\pi}{3}(\rho+3p+2P) ,
\label{fl2}
\ee
where $P$ is the pressure associated to the cosmological constant as $\Lambda= - 8\pi P$. This is an interesting analogy that allows us to treat the cosmological constant as a thermodynamic variable, which is effectively interpreted as a pressure. Then, from the expression for the FLRW apparent horizon (\ref{hor}), we obtain 
\begin{equation}
R_h=\sqrt{ \frac{3}{8\pi(\rho -P)}}\ ,
\label{rad2}
\end{equation}
which determines the horizon radius for $\rho > P$. 
Again,  the apparent horizon represents a causal horizon that can be interpreted as a thermodynamic system with fundamental equation
\begin{equation}
S= \pi R_h^2 = \frac{3}{8 (\rho-P)}\ . 
\label{ent2}
\end{equation}

\subsection{The first law}
\label{sec:flawl}

It has been argued that the presence of the cosmological constant affects the physical significance of the quantities entering the model \cite{dolan2010cosmological, kastor2009enthalpy, kubizvnak2017black, mann2016chemistry}.

\begin{figure}
	\includegraphics[scale=0.35]{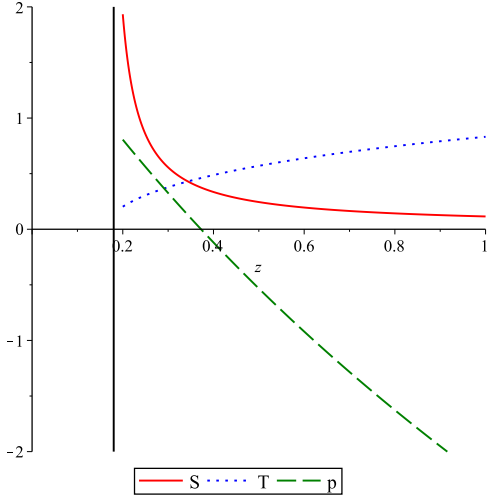}
	\caption{Entropy, temperature, and pressure in terms of the redshift.The constants are chosen as $a_0=1$, $P=1$, and $\rho_0=0.1$. The vertical solid line $z=0.18$ denotes the limit of existence of the apparent horizon.	
 }
	\label{fig2a}
\end{figure}

 In particular,  $\rho$ can be interpreted as the enthalpy density, i.e., 
\be
\rho = \frac{\cal H}{V}\ , \quad 
{\cal H} = E + PV\ .
\ee
Then,
\be
d{\cal H} = dE + PdV + VdP = - TdS + (W+P)dV + VdP\ ,
\label{dH1}
\ee
where we have used the first law (\ref{flaw}). 
Furthermore,  by using  Eq.(\ref{ent2}), we can compute the enthalpy and obtain  
\be
{\cal H }= \rho V = \frac{3}{8} \frac{V}{S} + PV\ ,
\ee
which leads to 
\be
d{\cal H} = - \frac{3}{8} \frac{V}{S^2} dS +\left(\frac{3}{8S}+P\right)dV + VdP\ .
\label{dH2}
\ee
Moreover,  a comparison of Eqs.(\ref{dH1}) and (\ref{dH2}) leads to 
\be
T= \frac{3}{8} \frac{V}{S^2}= \frac{8}{3}V(\rho-P)^2\ , \quad W = \frac{3}{8S} = \rho - P\ .
\label{temp3}
\ee
Furthermore, using the expression $W = \frac{1}{2}(\rho-p)$ that follows from the unified first law for the FLRW apparent horizon (\ref{uflaw2}), we obtain the equation of state
\be
\rho + p = 2P\ .
\label{eos2}
\ee
It follows that the presence of the cosmological constant  imposes a particular equation of state, which represents a linear relationship between pressure and density. This equation of state allows us to integrate the conservation law (\ref{claw}). The solution can be written out by using the definition of the redshift parameter as follows
\be
\rho = \rho_0 +  6 P \ln \frac{1+z}{a_0}\ ,
\label{dens2}
\ee
where $\rho_0$ is an integration constant. 
Consequently, in this case, the main thermodynamic variables evolve as 
\bea
S &=&  \frac{3}{8\left(\rho_0 - P +6P \ln \frac{1+z}{a_0} \right)}\ , \\
 T &= &\left(\frac{2}{3\pi}\right)^{1/2}    \left(\rho_0 - P + 6P\ln \frac{1+z}{a_0}\right)^{1/2} \ , \\
p &=& 2P -\rho_0 - 6 P \ln \frac{1+z}{a_0}
\ .
\eea
In Fig.\ref{fig2a}, we illustrate the dynamic behavior of these quantities. The apparent horizon exists only in the interval $\rho>P$, which corresponds to $z>\exp((1-\rho_0/P)/6)-1$ with $z> 0.18$ for $\rho_0=0$. The entropy function increases during the evolution as the density diminishes. The pressure is negative during most of the evolution and becomes positive as the lower limit of $z$ is approached.

\subsection{The second law}
\label{sec:slawl}

In this section, we analyze the consequences of demanding the fulfillment of the entropy conditions $\dot S\geq 0$ and $\ddot S\leq 0$. From the fundamental equation (\ref{ent2}), we obtain that 
\be
\dot S = - \frac{3}{8} \frac{\dot \rho}{(\rho-P)^2}\geq 0\ ,
\ee
which implies that the density should be a decreasing  function of time, i.e., $\dot \rho \leq 0$. According to the conservation law (\ref{claw}) and the equation of state (\ref{eos2}), this implies that the derivative of the scale factor should be a decreasing function of time, i.e., the velocity of expansion should decrease as the universe evolves.

As for the second entropy condition, from the fundamental equation (\ref{ent2}) and by using the conservation law (\ref{claw}), the Friedmann equations in the representation  (\ref{fl1}) and (\ref{fl2}), and the equation of state (\ref{eos2}), we obtain
\bea
\ddot S &=& \frac{3}{8(\rho-P)^2}\left(\frac{2\dot\rho^2}{\rho - P} - \ddot \rho\right) \\
&=& \frac{54 P^2}{(\rho-P)^2} - \frac{9k}{4} \frac{P(13P-\rho)}{a^2(\rho-P)^3} \leq 0 \ .
\eea
We see that in the case $k=0$, the condition $\ddot S\leq 0$ is satisfied only for $P=0$, which leads to the case analyzed in the last section. Consequently, from a thermodynamic point of view, a cosmological constant is not compatible with a spatially flat universe.  
 In universes with non-zero spatial curvature, the second entropy condition implies that $k(13P-\rho)>0$, indicating that the presence of the cosmological constant imposes limits on the values of $\rho$. Thus, for $k=1$ the enthalpy density is restricted to the range $\rho\in (P,13P)$, whereas for $k=-1$ it should be $\rho>13P$.
 
\begin{figure}
 	\includegraphics[scale=0.35]{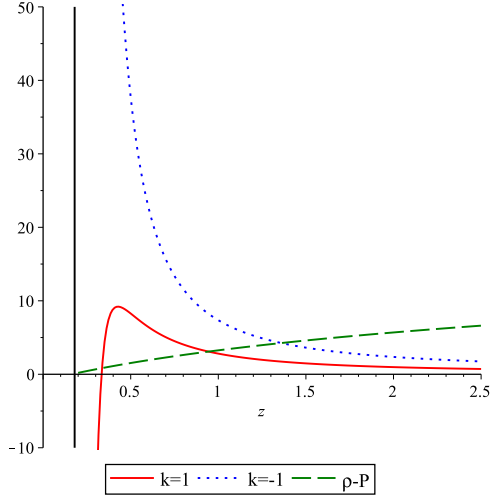}
 	\caption{Second derivative of the entropy as a function of the redshift for different values of the spatial curvature $k$. The constants are chosen as $a_0=1$, $P=1$, and $\rho_0=0.1$. The curve $\rho-P$ is also plotted. The vertical solid line indicates the lower bound for the existence of an apparent horizon.}
 	\label{fig3}
 \end{figure}

 In Fig. \ref{fig3}, we show the behavior of $\ddot S$ for different values of the parameters. First, the curve $\rho-P$ is plotted to determine the range of values of $z$, where the condition  $\rho>P$ is satisfied, i.e., the range where an apparent horizon can exist [cf. Eq.(\ref{rad2})]. For the chosen values of the parameters $a_0=1$, $P=1$, and $\rho_0=0.1$, this corresponds to the range $z>0.18$. Furthermore, the curve with $k=-1$ has only positive values in this range. Accordingly, in such a case, the apparent horizon cannot be interpreted as a thermodynamic system. The curve with $k=1$ is only partially contained in the region of negative values. In fact, one can see that the condition $\ddot S<0$ is fulfilled only within the range $z\in (0.13,0.33)$. Consequently, only within this interval the apparent horizon with cosmological constant can be interpreted as a perfect fluid with equation of state (\ref{eos2}).

 \subsection{The response functions}
 \label{sec:respl}
 
 To compute the response functions, we use the definitions given in Sec. \ref{sec:resp} with the relationships
 \be
 S = \sqrt{\frac{3V}{8T}}\ , \quad 
 V = \frac{3}{8} \frac{T}{(p-P)^2}\ ,
 \ee
 which follow from the fundamental equation (\ref{ent2}) and the expression for the temperature given in Eq.(\ref{temp3}). Then, a straightforward calculation leads to the following results
 \be
 C_V = -\frac{1}{2}\sqrt{\frac{3V}{8T}} \ , \quad 
 \kappa_T = -\frac{3}{4} \frac{T}{V(P-p)^3} \ , \quad
 \alpha = \frac{3}{8V(P-p)^2}\ ,
 \label{resp3}
 \ee
 with the adiabatic index
 \be
 \gamma = \frac{C_p}{C_V}= \frac{\kappa_T}{\kappa_S} = 2 \ .
 \ee
 We see that the cosmological constant appears explicitly only in the compressibility $\kappa_T$ and the thermal factor $\alpha$. It seems to induce a divergency when $p=P$. However, according to the equation of state (\ref{eos2}), this is equivalent to the condition $\rho=P$, a value which  is not allowed by the sole definition of the radius of the apparent horizon (\ref{rad2}). The heat capacity is negative as in the case of systems with long-range interactions.  
  Moreover, the adiabatic index is the same as in the case without cosmological constant and leads to the conclusion that the apparent horizon as a thermodynamic system has only two degrees of freedom ($f=2$). Again, we believe that this reduction in the number of degrees of freedom, when compared with the simplest monoatomic gas,  is due to the presence of a long-range interaction. 
 
 Finally, let us explore the dependence of the response functions in terms of the redshift. From Eqs.(\ref{resp3}), (\ref{temp3}), and (\ref{eos2}), we obtain that 
 \be
 C_V = - \frac{3}{16(\rho-P)}\ , \quad
 \kappa_T = - \frac{2}{\rho-P}\ , \quad 
 \alpha= \frac{3}{8V(\rho-P)^2}\ ,
 \ee
 which, by means of the density function (\ref{dens2}), depend all on the redshift. This means that in this case all the response functions are dynamic variables, which represents the main difference in comparison with the former case without cosmological constant. 
 
 We conclude that the apparent horizon in a FLRW spacetime with cosmological constant can be interpreted in certain time intervals as a thermodynamic system, which corresponds to a perfect fluid with a linear equation of state and time-dependent response functions.

\section{Conclusions} 
\label{sec:con}

In this work, we explored the properties of the apparent horizon of the FLRW spacetime from a thermodynamic point of view. It is by now well known that there exists a deep relationship between  the field equations of a homogeneous and isotropic spacetime and the first law of thermodynamics of the corresponding apparent horizon, once it assumed that the area of the horizon is related to its  entropy. To this end, it is necessary to assume that the horizon possesses certain thermodynamic properties. In this work, we go further in this direction and assume that the apparent horizon is in fact a thermodynamic system that can be explored with the standard laws of classical thermodynamics, where all the properties of the system cam be derived from the corresponding fundamental equation \cite{callen1998thermodynamics}. 

Thus, the main step towards the analysis of the apparent horizon as a thermodynamic systems consists in determining the corresponding fundamental equation. To this end, we use the conceptual basis of black hole thermodynamics, in which the fundamental equation is the entropy of the horizon. Therefore, we calculate the entropy of the apparent horizon of a FLRW spacetime in terms of the horizon radius. Then, using Friedmann equations, we rewrite the entropy in terms of thermodynamic variables and the result is interpreted as the fundamental equation of the dynamic apparent horizon.

First, we analyze the case of a FLRW spacetime with a barotropic perfect fluid as the gravitational source. We use the first law of thermodynamics to derive explicit expressions for the  main thermodynamic variables of the system. Moreover, from the unified first law for  apparent horizons, it follows that the perfect fluid must obey an equation of state equivalent to that of the dark energy. 
We also investigate the dynamic behavior of the entropy by demanding that it is an increasing function, which satisfies the principle of maximum entropy. It turns out that these conditions are satisfied only by certain values of the barotropic factor that depend on the type of spatial curvature of the cosmological model. Moreover, the second entropy condition ($\ddot S\leq 0$) can be violated during the evolution of the horizon, indicating that the interpretation as a thermodynamic system is no more valid. 

We also analyzed the behavior of the response functions of the apparent horizon. First, we notice that the heat capacities are always negative, a behavior that we interpret as due to the long-range character of the underlying gravitational interaction. Moreover, the heat capacities, compressibilities and the adiabatic index as well  turn out to be constant during the entire evolution, a behavior that resembles that of ideal gases. This allows us to associate the adiabatic index with the number of degrees of freedom ($f$) of the system, which turns out to be equal to 2. This number is less than the number of degrees of freedom of a simple monoatomic ideal gas for which $f=3$. 

We conclude that the FLRW apparent horizon can be interpreted as thermodynamic system corresponding to  a ``dark-energy" fluid, with negative heat capacities and compressibilities and a low number of degrees of freedom.

In the case of FLRW spacetime with cosmological constant the situation is different. Firs of all, the first law of thermodynamics implies that the fluid must satisfy a linear equation of state, in which the cosmological constant appears explicitly in the intercept of the function. As a result, all the thermodynamic variables change during the evolution of the horizon. Moreover, the presence of the  cosmological constant imposes limits on the values of the enthalpy density, implying that  apparent horizons can be interpreted as thermodynamic systems only for certain values of the cosmological constant, which depend also on the value of the spatial curvature of the FLRW model. In particular, in the case of spatially flat universes (with non-zero cosmological constant), no apparent horizon can be interpreted as a thermodynamic system. We also found that in this case the heat capacities and compressibilities are negative quantities, and the system has only two degrees of freedom, results that interpret again as due to the long-range character of the gravitational interaction.

In conclusion, we can say that the analogy between gravity and thermodynamics can also be applied to apparent horizons of cosmological models. However, some care is needed when interpreting apparent horizons as thermodynamic systems because the analogy is not always valid for all the parameters that enter the dynamics of the horizon and certainly not for the entire evolution of the horizon.  

\section*{Acknowledgments}

This work was partially supported  by UNAM-DGAPA-PAPIIT, Grant No. 114520, and Conacyt-Mexico, Grant No. A1-S-31269.


%

\end{document}